\documentclass{article}
\usepackage{graphicx} 
\usepackage{subcaption} 
\usepackage{amsmath} 
\usepackage{geometry} 
\usepackage{setspace} 
\onehalfspace
\usepackage{enumitem} 
\usepackage{authblk} 
\usepackage[bottom]{footmisc} 
\usepackage{multicol}

\usepackage[
backend=biber,
style=chem-acs,
sorting=none
]{biblatex}
\title{\textbf{Guidelines for the optimization of hafnia-based ferroelectrics through superlattice engineering}\\[1ex]}

\begin{document}

\renewcommand\Affilfont{\itshape\small}
\renewcommand{\thefootnote}{\fnsymbol{footnote}}

\author[1]{Johanna van Gent\footnote{j.van.gent.gonzalez@rug.nl}} 
\author[2]{Binayak Mukherjee\footnote{binayak.mukherjee@list.lu}}
\author[1]{Ewout van der Veer\footnote{ewout.van.der.veer@rug.nl}}
\author[3]{Ellen M. Kiens\footnote{e.m.kiens@utwente.nl}}
\author[3]{Gertjan Koster\footnote{g.koster@utwente.nl}}
\author[1]{Bart J. Kooi\footnote{b.j.kooi@rug.nl}}
\author[2,4]{Jorge Íñiguez-González\footnote{jorge.iniguez@list.lu}}
\author[1]{Beatriz Noheda\footnote{b.noheda@rug.nl}} 

\affil[1]{Zernike Institute for Advanced Materials, University of Groningen, Groningen, The Netherlands}
\affil[2]{Smart Materials Unit, Luxembourg Institute for Science and Technology (LIST), Esch-sur-Alzette, Luxembourg}
\affil[3]{MESA+ Institute for Nanotechnology, University of Twente, Enschede, The Netherlands}
\affil[4]{Department of Physics and Materials Science, University of Luxembourg, Belvaux, Luxembourg}

\date{}

\maketitle

\renewcommand\thesection{\arabic{section}}
\captionsetup[figure]{labelfont={bf},labelformat={default},labelsep=period,name={Figure}}

\begin{figure}[ht]
    \renewcommand{\figurename}{Figure}
    \renewcommand{\thefigure}{}
    \centering
    \includegraphics[width=0.75\textwidth]{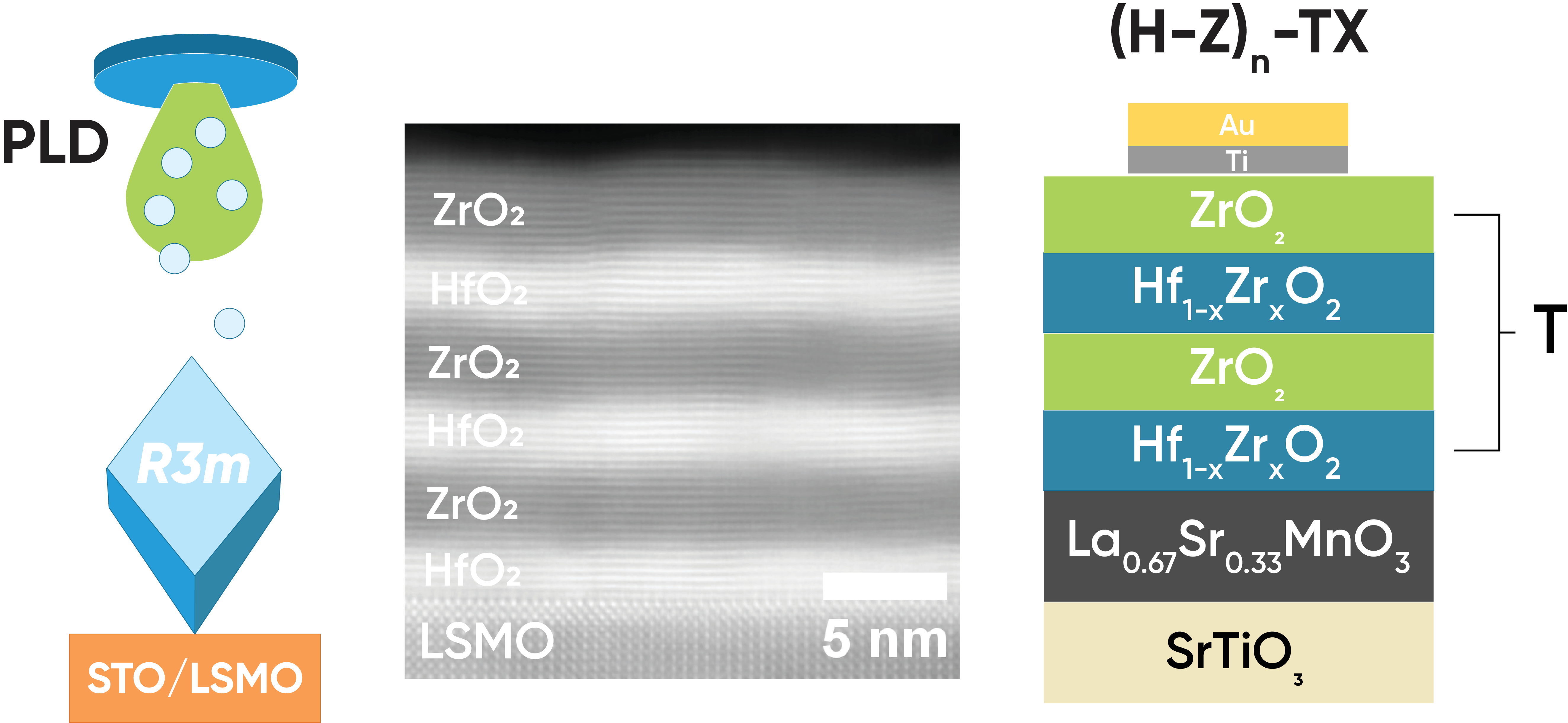}
     \hfill
\end{figure}

\pagebreak

\section{Abstract}

Hafnia-based ferroelectrics are revolutionizing the data storage industry and the field of ferroelectrics, with improved materials and devices being reported monthly. However, full understanding and control has not been reached yet and the ideal material still needs to be found. Here we report ferroelectric hafnia-zirconia superlattices made out of  zirconium-substituted hafnia (Hf$_{1-x}$Zr$_x$O$_2$) sublayers of varying stoichiometries alternating with pure ZrO$_2$ sublayers. It is observed that the ZrO$_2$ layers in these superlattices act as a booster for the total remnant polarization (P$_r$). By combining the benefits of the ZrO$_2$ layers and the added interfaces, which help prevent breakdown, we fabricate  superlattices with a total 87.5\% ZrO$_2$ content, exhibiting record polarizations with a 2P$_r$ value of 84 µC/cm$^2$ that can be cycled 10$^9$ times, while maintaining a 2P$_r$ $>$ 20 $\mu$ C/cm$^2$2. Next to these attractive properties, substitution of HfO$_2$ by the much more abundant ZrO$_2$ offers a significant step towards the sustainable application of these devices. 

\pagebreak

\section{Introduction}
Ever since the concepts of ferroelectric capacitors as non-volatile memory were birthed in the 1950s \cite{Buck1952FerroelectricsSwitching,Ross1957SemiconductiveDevice,Brown1957SemiconductiveDevice, Looney1957SemiconductiveDevice,Morton1957ElectricalStorage}, effort has been made in fulfilling the promise of such devices for faster, more efficient data storage hardware \cite{Scott1989FerroelectricMemories,Auciello1998TheMemories}. Ferroelectric devices introduce a ferroelectric thin film to established device architectures, where the addition of a switchable spontaneous polarization enables storage of binary values as up/down polarization within the ferroelectric layer. However, the prospect for ferroelectric devices was initially limited since the perovskite-based complex oxide ferroelectrics that dominated the most prominent era in ferroelectric device research of the 80s-00s, depolarize in sub-10 nm thicknesses and cannot be easily integrated into existing CMOS architectures.\cite{Muller2015FerroelectricProspects,Salahuddin2018TheElectronics,Schenk2020MemoryScientists}

In 2011, the first report of robust ferroelectricity in silicon-doped HfO\textsubscript{2} \cite{Boscke2011FerroelectricityFilms} was initially met with skepticism: ferroelectric HfO\textsubscript{2}  was shown to overcome not one but both major limitations of other ferroelectric materials: i) it did not depolarize at the nanoscale and ii) its history as a high-permittivity (high-k) gate insulator for high-performance field-effect transistors (FETs) proved it was compatible with CMOS technology. Over time this behaviour was proven to be reproducible, leading to gradual acceptance and a subsequent explosion of research into ferroelectric hafnia. The origin of ferroelectric behaviour in these first polycrystalline films synthesized by atomic layer deposition (ALD) was attributed to a metastable orthorhombic phase (\(Pca2_1, OIII\)), but the mechanisms that make this material so different from other ferroelectrics (\emph{i.e.} the reason why it does not depolarize at the nanoscale) remains unknown \cite{Boescke2011FerroelectricityTransistors, Schenk2020MemoryScientists,Noheda2023LessonsFerroelectrics}.

Since then, extensive research into the material has improved understanding regarding optimization of ferroelectric performance in HfO\textsubscript{2}-based devices. The principal hurdle in obtaining ferroelectric hafnia is that the non-polar monoclinic \(P2_1/c, \ m\) phase is far more stable than the metastable polar \(OIII\) phase. Consequently, any strategy for achieving the metastable ferroelectric phase hinges on avoiding the large unit cell volume of the former phase. As the window of stability of the polar phase is small, the easiest approach is to first stabilize the much lower volume tetragonal \(P4_{2}/nmc\) phase and then slightly enlarge the cell such that it may form the polar phase. There are several mechanisms capable of achieving this, which are often used simultaneously for enhanced stability and performance. 

Firstly (and most prominently), the use of surface tension through synthesis of small grains in polycrystalline films or thin single crystal films forces a smaller unit cell volume, thereby stabilizing the tetragonal or even cubic (\(Fm\bar3m\)) phase at annealing temperatures which, upon cooling, relax into the metastable \(OIII\) phase \cite{Materlik2015TheModel,Chae2020StabilityFilms}. In the case of epitaxial thin films, the strain associated with the surface tension is further aided by the additional substrate-induced (epitaxial) strain which will also dictate the orientation of the polar axis \cite{Shimizu2016TheFilm,Katayama2016OrientationFilms,Estandia2019EngineeringStress,Nukala2019DirectSilicon}. 

In addition, doping HfO\textsubscript{2} with various elements (Si, Zr, La, Y, etc.) has also been shown to be an effective strategy in improving ferroelectric phase stability and device performance for both polycrystalline and epitaxial thin films through the associated "chemical pressure" induced by dopants \cite{Starschich2017AnHfO2,Park2019DopantsFilms,Tashiro2021ComprehensiveFilms,Song2022SynergeticFilms,Yun2022IntrinsicFilms}. A popular choice of hafnia composition is the Hf\textsubscript{0.5}Zr\textsubscript{0.5}O\textsubscript{2} solid solution, which serves to both stabilize the ferroelectric phase and to increase the remnant polarization relative to pure hafnia \cite{Materlik2015TheModel}. Similarly, the strain associated with the presence of oxygen vacancies has also been shown to be beneficial in both polycrystalline and epitaxial thin films \cite{Jaszewski2022ImpactSputtering,Kaiser2023CrystalOxide}.

There is one important distinction between polycrystalline and epitaxial hafnia films: since epitaxial films enable control of the crystalline direction, the polar axis of the film is uniform across the layer and there is no need to "wake-up" the films by electrical cycling after growth \cite{Wei2018AFilms}. An additional distinguishing factor is that there are two different metastable ferroelectric phases accessible to epitaxial thin films: the \(OIII\) phase mentioned thus far \cite{Shimizu2016TheFilm,Estandia2019EngineeringStress,Yun2022IntrinsicFilms} and two rhombohedral phases that display similar behaviour. These non-centrosymetric rhombohedral \(R3m\) \cite{Wei2018AFilms,Kaiser2023CrystalOxide} and \(R3\) \cite{Nukala2020GuidelinesFilms,Begon-Lours2020StabilizationFilms,Guo2024RhombohedralFerroelectricity} phases have previously been found in epitaxial films under high compressive strains and exhibit large polarization values. However, they typically suffer from larger coercive fields, close to the breakdown fields, which increase the operation power and pose a challenge to the device cyclability and endurance.

In the particular case of epitaxial films grown on SrTiO\textsubscript{3}, the choice of bottom electrode has been found to be vital for \(R3m\) phase stability through an as-of-yet unclear mechanism. It seems thus far that only La\textsubscript{1-x}Sr\textsubscript{x}MnO\textsubscript{3} enables growth of high quality r-phase epitaxial hafnia films, which has been hypothesized to be a result of either its role as an oxygen reservoir, or as buffer layer providing epitaxial strain relaxation or due to charge transfer \cite{Nukala2021ReversibleDevices,Shi2023Interface-engineeredFilms}.

Recently, there has been increased interest in the synthesis of HfO\textsubscript{2}-ZrO\textsubscript{2} multilayers, which have been demonstrated to enhance spontaneous polarizations and cyclability relative to equivalent solid solution Hf\textsubscript{1-x}Zr\textsubscript{x}O\textsubscript{2} for ALD multilayers \cite{Weeks2017EngineeringNanolaminates,Gong2022PhysicalFilm,Lehninger2023FerroelectricReliability,Begon-Lours2024Back-End-of-LineNanolaminates}. In the context of epitaxial superlattices, only one report of Hf$_{0.5}$Zr$_{0.5}$-HfO$_2$ superlattices grown on SrTiO$_3$ exists, where an increase to both the polarization and dielectric response was observed at the cost of device cyclability \cite{Farahani2025DualNanolaminates}. The latter observation is surprising as increased cyclability has been readily observed in ALD multilayers, where it has been attributed to redistribution of defects along the multilayer interfaces \cite{Gong2022PhysicalFilm}, which can still take place in epitaxial superlattices and implies the observed poor cyclability may be due to other conflicting effects such as pinned defects. Note that here a distinction is made between multilayers, which refer to any stacking of different materials without a well-defined crystallographic ordering, and superlattices, which involve the presence of periodic crystallographic order that may be defined in terms of the unit cells of its constituent sublayers.

Ideally, one could expect that a superlattice construction addresses the limited cyclability of rhombohedral epitaxial hafnia films while harnessing their large remanent polarization and enhancing r-phase stability, due to the symmetry breaking inherent to superlattice geometries, as well as the strain brought about by additional interfaces. Furthermore, it can be expected that for superlattices based on r-phase hafnia (grown on SrTiO\textsubscript{3}(001) with a La\textsubscript{0.67}Sr\textsubscript{0.33}MnO\textsubscript{3} buffer) the r-phase of ZrO\textsubscript{2} is readily obtained since it has been shown that ferroelectric r-phase ZrO\textsubscript{2} can be obtained due to size effects (stabilized by surface energy contributions) for thicknesses below 33 nm \cite{ElBoutaybi2022StabilizationEnergy}. In fact, it is expected that the formation of fully ferroelectric r-phase HfO\textsubscript{2}-ZrO\textsubscript{2} occurs more readily than the \emph{OIII} equivalents, as \emph{OIII}-phase ZrO\textsubscript{2} cannot be stabilized only by size effects \cite{Materlik2015TheModel} and has only been reported for polycrystalline films below 7 nm \cite{Huang2021Sub-7-nmFerroelectricity,Cheema2022EmergentSilicon}.

In this work, HfO\textsubscript{2}- ZrO\textsubscript{2} superlattices are grown epitaxially for the first time. Hf\textsubscript{1-x}Zr\textsubscript{x}O\textsubscript{2}- ZrO\textsubscript{2} superlattices are grown on SrTiO\textsubscript{3}(001)/La\textsubscript{0.67}Sr\textsubscript{0.33}MnO\textsubscript{3} by pulsed laser deposition (PLD). Comparison to solid solution films reveals that the appropriate choice of a superlattice provides a notable improvement to both remnant polarization (P\textsubscript{r}) and device cyclability, reaching values that surpass those of ALD nanolaminates reported thus far. It is found that the polarization is mostly influenced by the Hf\textsubscript{1-x}Zr\textsubscript{x}O\textsubscript{2} sublayer stoichiometry and total stack thickness, while the cyclability is mostly dependent on the number of interfaces present in the superlattice stack.

\section{Results \& Discussion}
\setcounter{figure}{0}

\subsection{Stabilizing rhombohedral Hf\textsubscript{1-x}Zr\textsubscript{x}O\textsubscript{2}-ZrO\textsubscript{2} superlattices}
    
\begin{figure}[ht!]
    \centering
        \includegraphics[width=1\textwidth]{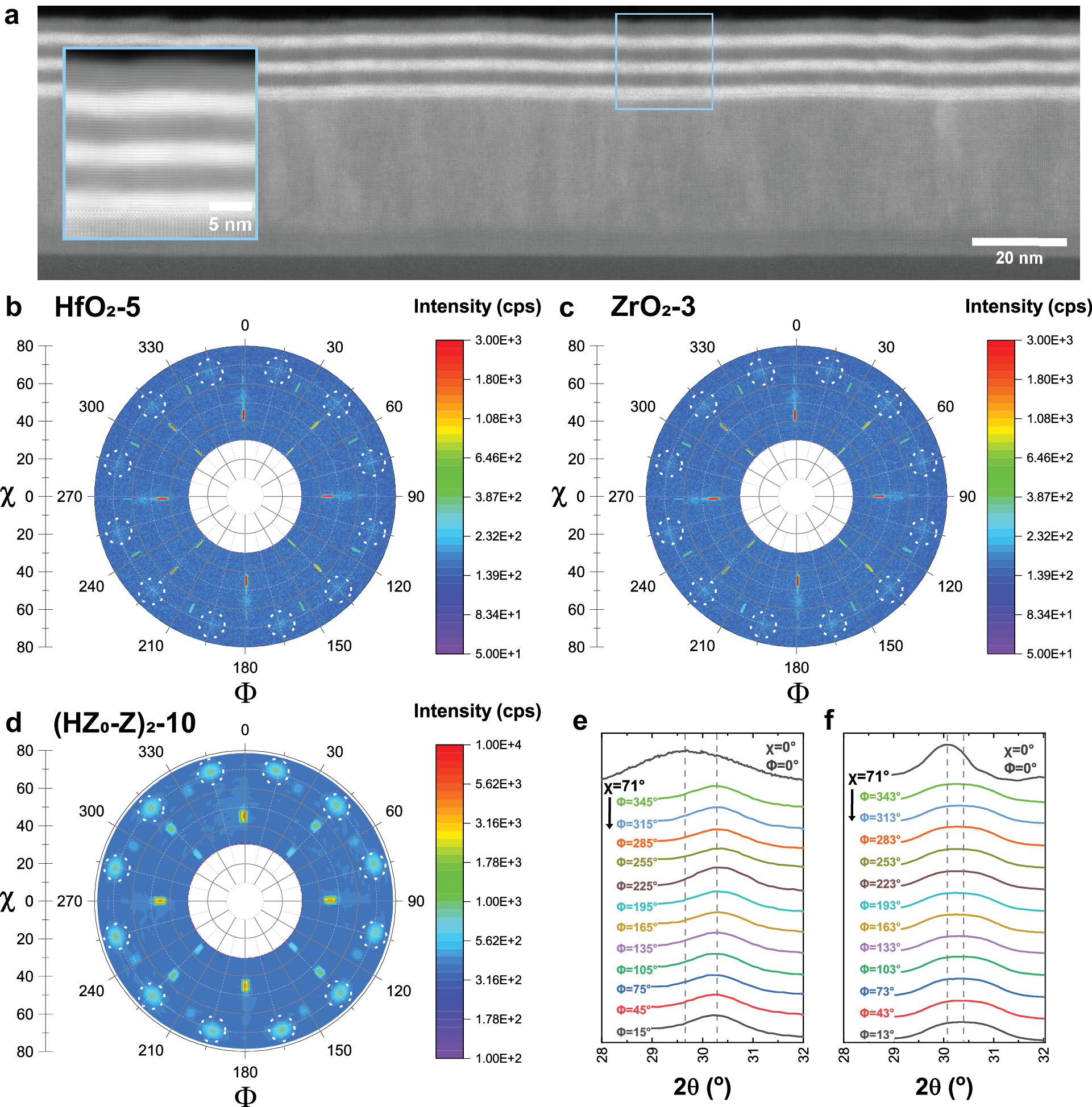}
        \hfill
    \caption{The STEM-HAADF overview image in \emph{a)} demonstrates that the interfaces are atomically sharp, with low roughness and negligible ion interdiffusion. A strong out-of-plane texture exists within the superlattices (see also inset). This figure also shows that good sublayer thickness control (in this case 2.5 nm) has been attained. Since in projection (in the TEM lamella produced by FIB) overlapping domains with different in-plane rotations are probed simultaneously, STEM is not able to provide a clear view of the superlattice crystal structure and, as a result, further structural characterization will rely on various x-ray diffraction measurements. Pole figures around the (111) out-of-plane peak seen in Figure \ref{fig:XRDComp} for \emph{b)} HfO\textsubscript{2}-5 and \emph{c)} ZrO\textsubscript{2}-3 monolayers, as well as for a \emph{d)} (HZ\textsubscript{0}-Z)\textsubscript{2}-10 superlattice. The white dotted circles overlayed over the pole figures indicate the positions of the poles corresponding to the Hf\textsubscript{1-x}Zr\textsubscript{x}O\textsubscript{2} film. \emph{e)} and \emph{f)} $\theta-2\theta \ $ in-plane measurements around each of the poles observed in \emph{c)} and \emph{d)}, respectively.}
    \label{fig:PF_Comp}
\end{figure}

Superlattices with n repetitions of alternating sublayers of Hf\textsubscript{1-x}Zr\textsubscript{x}O\textsubscript{2} and ZrO\textsubscript{2} (HZ\textsubscript{x}-Z)\textsubscript{n} have been grown, as described in section 2. Figure \ref{fig:PF_Comp}a demonstrates that the interfaces are atomically sharp, with low roughness and negligible ion interdiffusion. It also shows that good sublayer thickness control (in this case 2.5 nm) has been attained. Due to the small domain size in comparison to the thickness of the TEM lamella overlapping domains are probed simultaneously, meaning no clear view of the superlattice crystal structure can be found by STEM and, as a result, further structural characterization will rely on various x-ray diffraction measurements.

\begin{figure}[ht!]
    \centering
    \includegraphics[width=0.99\textwidth]{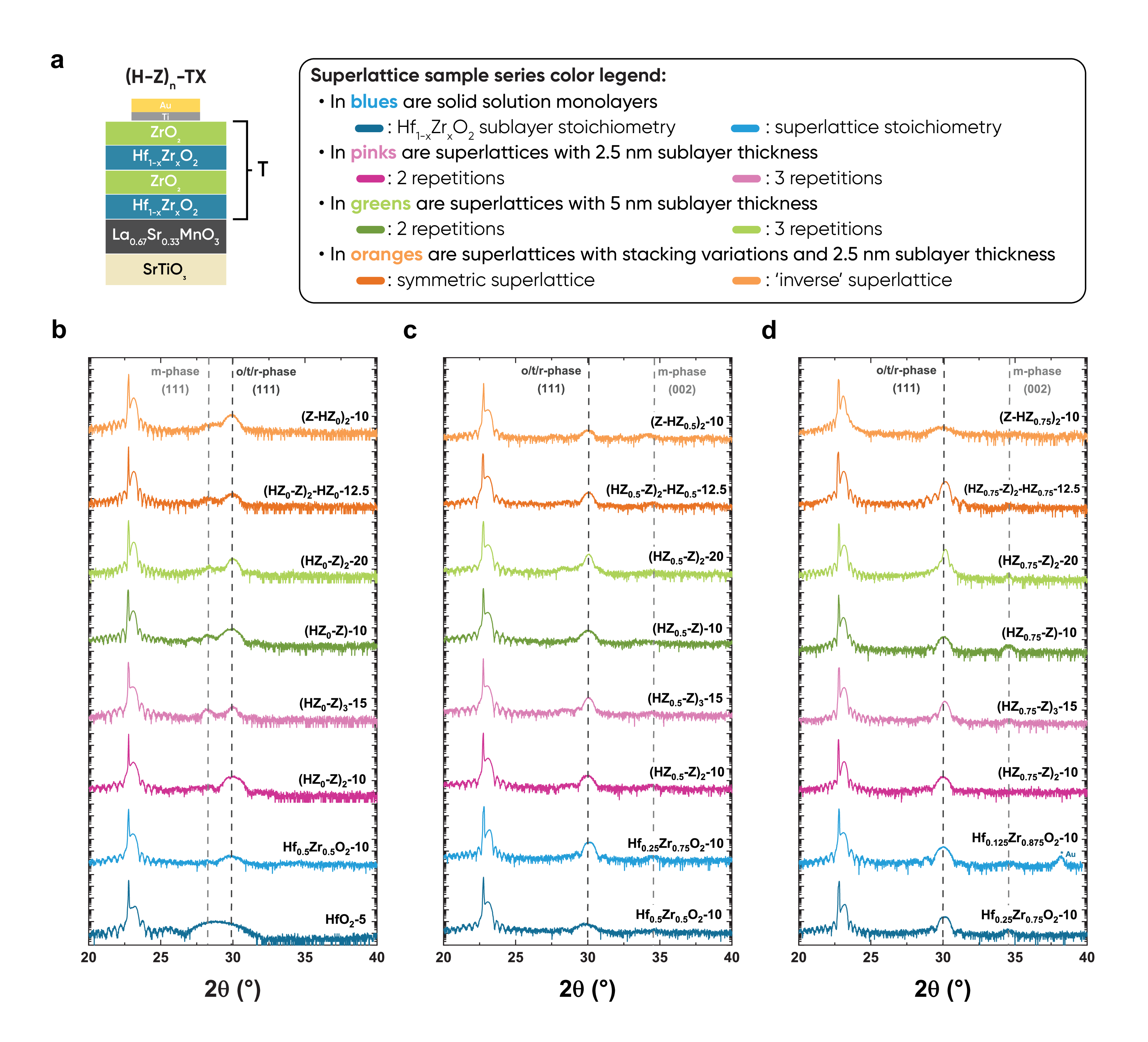}
     \hfill
        \caption{\emph{a)} Schematic diagram of the superlattice devices with a legend for the color coding of the series of superlattice samples. \emph{b-d)} $\theta-2\theta \ $ diffractograms of the three tested families of superlattices: \emph{b)} (HZ\textsubscript{0}-Z)\textsubscript{n}-T, \emph{c)} (HZ\textsubscript{0.5}-Z)\textsubscript{n}-T and \emph{d)} (HZ\textsubscript{0.75}-Z)\textsubscript{n}-T. The thickness (T) labelled on the diffractograms refers to the total thickness (in nm) of the sample (monolayer or superlattice stack). In the superlattices shown here, the thickness of  Hf\textsubscript{1-x}Zr\textsubscript{x}O\textsubscript{2} layers is the same as that of ZrO\textsubscript{2} layers.}
        \label{fig:XRDComp}
\end{figure}

Figure \ref{fig:XRDComp} shows $\theta-2\theta \ $ scans for several configurations of HZ$_x$-Z superlattices, containing either  a) HfO\textsubscript{2}, b) Hf\textsubscript{0.50}Zr\textsubscript{0.50}O\textsubscript{2} or c) Hf\textsubscript{0.25}Zr\textsubscript{0.75}O\textsubscript{2} as the HZ layer in the superlattice. The results of Figure \ref{fig:XRDComp} resemble other XRD measurements of epitaxial Hf\textsubscript{0.50}Zr\textsubscript{0.50}O\textsubscript{2} grown on LSMO-buffered ST0(001) with the peaks at 2$\theta$=22.8° and 2$\theta$=23.2° corresponding to the STO (001) and the LSMO (pseudocubic) (001) reflections, respectively \cite{Wei2018AFilms}. The most prominent hafnia/zirconia peaks appear at 2$\theta$=30° and may correspond to the (111) reflections of the rhombohedral, orthorhombic \emph{OIII} or tetragonal phases, where additional peaks in the 28-35° range are indicative of relaxation into the monoclinic phase.
As the \emph{OIII} and r-phases cannot be distinguished from such out-of-plane $\theta-2\theta \ $ measurements, pole figures around the (111) reflection of either phase (at $2\theta\approx30°$) \ are carried out. Figure \ref{fig:PF_Comp} shows the pole figures of one of the superlattices compared with the patterns obtained for a single HfO\textsubscript{2} layer and a single ZrO\textsubscript{2} layer of similar total thickness. In pure HfO\textsubscript{2}, a 12-fold symmetry is observed, where $\theta-2\theta \ $ scans around each of the observed poles show that the diffraction angle for the out-of-planes d\textsubscript{111} is offset by \textasciitilde0.6° from that of the in-plane set of \{111\} planes. As previously reported \cite{Wei2018AFilms}, such results can be accounted for only if the obtained phase is rhombohedral. The 12-fold multiplicity around the (111)-direction can then be ascribed to the presence of 4 domains of 3-fold symmetry. 

More notably, Figure \ref{fig:PF_Comp}c also suggests that pure ZrO\textsubscript{2} can be stabilized in the same non-centrosymmetric r-phase as HfO\textsubscript{2} on LSMO-buffered STO(001). However, this is only possible over a small range of film thicknesses (2-4 nm). At larger thicknesses, relaxation to the otherwise more stable tetragonal (\emph{P4\textsubscript{2}/nmc}) phase takes place. Although this may seem unexpected at first, the existence of a strain-induced rhombohedral ZrO\textsubscript{2} has been known for decades \cite{Hasegawa1983RhombohedralPSZ} and recent \emph{ab initio} calculations suggest that the stabilization of the r-phase by surface effects alone is easier in ZrO\textsubscript{2} than in HfO\textsubscript{2} \cite{ElBoutaybi2022StabilizationEnergy}. 

\begin{figure}[ht]
    \centering
    \includegraphics[width=1\textwidth]{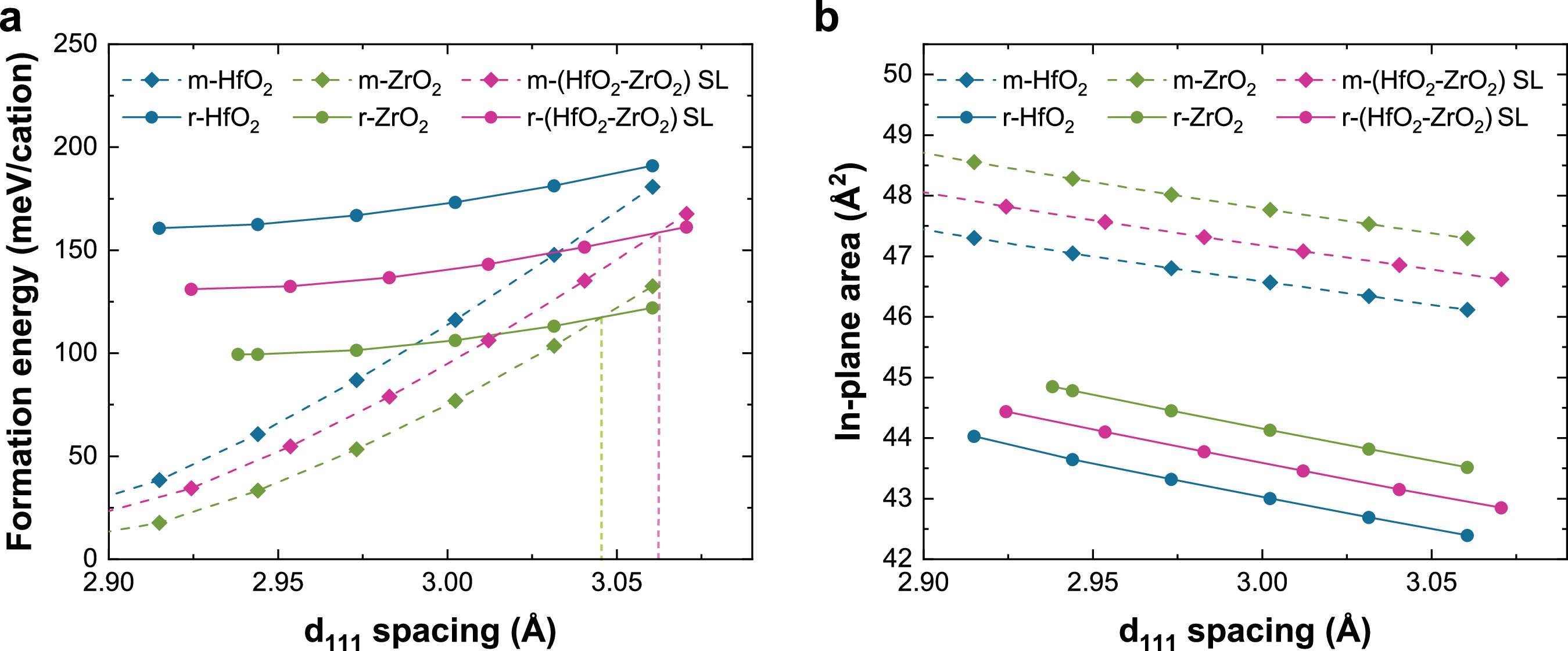}
     \hfill
        \caption{Comparison of \emph{a)} formation energies and \emph{b)} in-plane area versus d\textsubscript{111} for the monoclinic and rhombohedral phases of HfO\textsubscript{2}, HfO\textsubscript{2}-ZrO\textsubscript{2} superlattices and ZrO\textsubscript{2}. The \(d_{111}\) spacings for 3 nm HfO$_2$ and ZrO$_2$ layers, as characterized by XRD, are 3.14 Å and 3.01 Å, respectively.}
        \label{fig:DFT}
\end{figure}

The DFT results of Figure \ref{fig:DFT} show the formation energy of both the m-phase and r-phase as a function of the out-of-plane d$_{111}$ spacing, which is in turn related to the in-plane area, \emph{i.e.} to the in-plane strain. Figure \ref{fig:DFT}a demonstrates that, for ZrO$_2$, as well as for HfO$_2$-ZrO$_2$ superlattices, the r-phase can become more stable than the ground state m-phase through epitaxial strain alone and regardless of oxygen content, in agreement with a previous study \cite{Kaiser2023CrystalOxide}. These results further suggest that ZrO$_2$ more readily forms the r-phase under compressive strain than HfO$_2$, likely due to the smaller in-plane area compared to its m-phase. Furthermore, these results show that the formation energy for r-phase HfO\textsubscript{2}-ZrO\textsubscript{2} superlattices lies in the middle of that of its constituent HfO\textsubscript{2} and ZrO\textsubscript{2} layers.

Consequently, the thickness of ZrO\textsubscript{2} layers is fixed to <5 nm in all superlattices to ensure there is a minimal amount of t-phase present. It should be noted, however, that in superlattices with HfO\textsubscript{2} (which relaxes from the r- into the m-phase when thicker than 7 nm on STO(001)/LSMO, as shown in SI and is reminiscent of the relaxation of the Hf\textsubscript{0.5}Zr\textsubscript{0.5}O\textsubscript{2} composition in Wei \emph{et al.} \cite{Wei2018AFilms}) stability of polar r-ZrO\textsubscript{2} relative to antipolar t-ZrO\textsubscript{2} is expected to be further enhanced by proximity effects, that is by the electric field generated by the neighbouring ferroelectric r-HfO\textsubscript{2} layers \cite{Neaton2003TheorySuperlattices,Mukherjee2024First-principlesSuperlattices}. Indeed the pole figure and in-plane $\theta-2\theta \ $ measurements of Figure \ref{fig:PF_Comp}c for a (HZ\textsubscript{0}-Z)\textsubscript{2}-10 superlattice possesses the same characteristics of the r-HfO\textsubscript{2} and r-ZrO\textsubscript{2} films and no indication of additional phases is present, despite the presence of 5 nm ZrO\textsubscript{2} layers in the stack. 

Figures \ref{fig:XRDComp} and \ref{fig:PF_Comp} show that growth of ferroelectric (111)-oriented r-phase forms the dominant contribution for all three families of superlattices. In some cases, out-of-plane measurements (Figure \ref{fig:XRDComp}) show additional peaks at 2$\theta$=28° and 2$\theta$=34.5° corresponding to the (\(\bar111\)) and (\(020\)) reflections of the monoclinic phase, respectively, which are indicative of strain relaxation in the superlattice. Comparing the different samples, several conclusions can be made regarding r-phase stability in epitaxial Hf\textsubscript{1-x}Zr\textsubscript{x}O\textsubscript{2}-ZrO\textsubscript{2} superlattices:

\begin{enumerate}[label=\roman*]

\item The higher the Zr content of the Hf\textsubscript{1-x}Zr\textsubscript{x}O\textsubscript{2} layers, the less prone they are to thickness relaxation into m-Hf\textsubscript{1-x}Zr\textsubscript{x}O\textsubscript{2}, thus requiring less strain for stabilization over the ground state m-phase.

\item Superlattices with thinner ZrO\textsubscript{2} and Hf\textsubscript{1-x}Zr\textsubscript{x}O\textsubscript{2} sublayers show less thickness relaxation than those with thicker sublayers but equivalent total thickness (see e.g. how (HZ\textsubscript{x}-Z)\textsubscript{2}-10 compares with (HZ\textsubscript{x}-Z)-10). For superlattices with sufficiently thin sublayers, it is expected that the most energetically favourable way of compensating large depolarizing fields caused by an out-of-plane polar discontinuity is simply to adopt a uniform polarization throughout the stack \cite{Neaton2003TheorySuperlattices, Dawber2005UnusualSuperlattices, Aguado-Puente2011InterplaySuperlattices, Zubko2012ElectrostaticSuperlattices}. Because maintaining a uniform polarization becomes increasingly energetically costly with increasing sublayer thickness, rhombohedral phase stability in both ZrO\textsubscript{2} and Hf\textsubscript{1-x}Zr\textsubscript{x}O\textsubscript{2} sublayers is enhanced if these are thin.

\item Although within the family of superlattices, the number of interfaces help to stabilize the r-phase, comparison of a superlattice with a solid solution film of equivalent stoichiometry and thickness shows that superlattices are not inherently better at stabilizing the r-phase than solid solutions. Rather, it appears that the most effective method for r-phase stabilization lies in maximizing the Zr-content through the Hf\textsubscript{1-x}Zr\textsubscript{x}O\textsubscript{2} sublayers (increasing the thickness of the ZrO\textsubscript{2} sublayers is ineffective as these relax to t-phase).

\item Superlattices starting with a ZrO\textsubscript{2} layer ("inverse", (Z-HZ)\textsubscript{n}) have larger fractions of t-ZrO\textsubscript{2} and m-HfO\textsubscript{2} than those that begin with a  Hf\textsubscript{1-x}Zr\textsubscript{x}O\textsubscript{2} layer ((HZ$_x$-Z)\textsubscript{n}). It is possible that the necessary interface with LSMO (previously shown to be an important factor for growth of ferroelectric HZO on STO \cite{Wei2018AFilms,Shi2023Interface-engineeredFilms}) may not form in the absence of Hf and that, in the absence of an optimal polar starting layer, stabilization of the rhombohedral phase is hindered.

\end{enumerate}

\subsection{Boosting P\textsubscript{r} and cyclability in Hf\textsubscript{1-x}Zr\textsubscript{x}O\textsubscript{2}-ZrO\textsubscript{2} superlattices}

\begin{figure}[ht]
    \centering
    \includegraphics[width=1\textwidth]{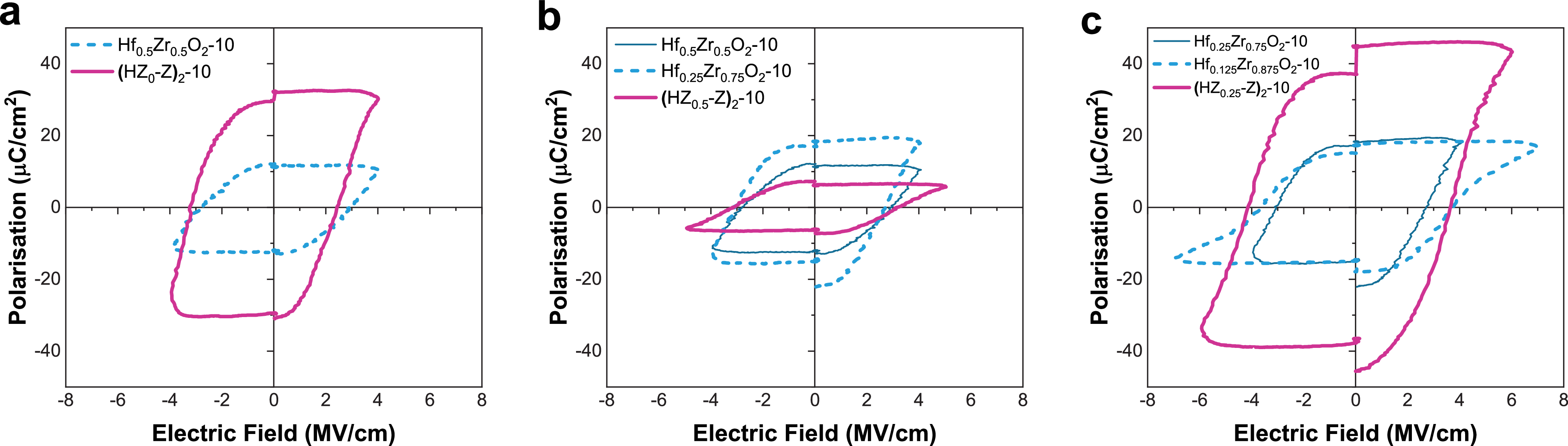}
     \hfill
        \caption{PUND ferroelectric measurements for the best performing superlattice of each of the three tested families (\emph{a)} (HZ\textsubscript{0}-Z)\textsubscript{n}-T, \emph{b)} (HZ\textsubscript{0.5}-Z)\textsubscript{n}-T and \emph{c)} (HZ\textsubscript{0.75}-Z)\textsubscript{n}-T (solid magenta lines). For comparison, in each panel, the loops are plotted together with those of two different Hf$_{1-x}$Zr$_x$O$_2$ monolayers of the same total thickness: one with the same composition as the Hf$_{1-x}$Zr$_x$O$_2$ sublayer (thin dark blue line) and a second one, with the same Hf/Zr ratio of the full superlattice (dashed light blue line). In a) the pure HfO\textsubscript{2} monolayer of equivalent thickness is not shown, since 10 nm layers show no ferroelectricity (a 5 nm layer displays P\textsubscript{r}=10 $\mu$ C/cm\textsuperscript{2} and E\textsubscript{c}=6 MV/cm, see SI).}
        \label{fig:PComp}
\end{figure}

\begin{figure}[ht]
    \centering
    \includegraphics[width=1\textwidth]{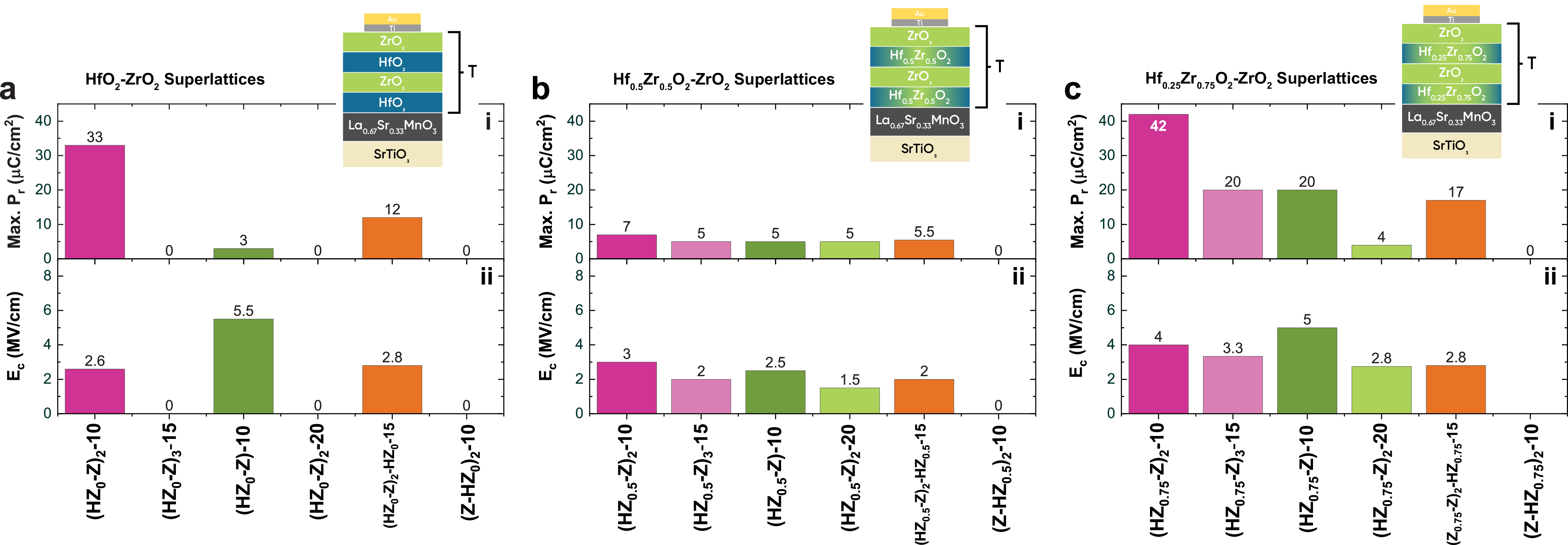}
     \hfill
        \caption{Comparison of \emph{i.} remnant polarization (P\textsubscript{r}, top) and \emph{ii.} coercive field (E\textsubscript{c}, bottom) values for each the three tested families of superlattices (\emph{a)} (HZ\textsubscript{0}-Z)\textsubscript{n}-T, \emph{b)} (HZ\textsubscript{0.5}-Z)\textsubscript{n}-T and \emph{c)} (HZ\textsubscript{0.75}-Z)\textsubscript{n}-T).}
        \label{fig:P&EComp}
\end{figure}

Having established the rules associated with maximising r-phase content in Hf\textsubscript{1-x}Zr\textsubscript{x}O\textsubscript{2}-ZrO\textsubscript{2} superlattices, their ferroelectric performance is evaluated. The PUND ferroelectric loops of three families of superlattices with the same total thickness of 10 nm, (HZ$_{0}$-Z)$_2$-T10, (HZ$_{0.5}$-Z)$_2$-T10 and (HZ$_{0.75}$-Z)$_2$-T10, that is with increasing ZrO$_2$ content in the Hf\textsubscript{1-x}Zr\textsubscript{x}O\textsubscript{2} sublayer, are plotted with solid lines in Figures \ref{fig:PComp} a, b and c, respectively. In each panel, the loops are compared with two types of Hf\textsubscript{1-x}Zr\textsubscript{x}O\textsubscript{2} monolayers of the same total thickness: one with the same composition as the Hf\textsubscript{1-x}Zr\textsubscript{x}O\textsubscript{2} sublayer and a second one, with the same Hf/Zr ratio of the full superlattice. In Figure \ref{fig:P&EComp}, the comparison of the remnant polarizations, P\textsubscript{r}, and coercive fields,  E\textsubscript{c}, is extended to other superlattices and represented as bar diagrams, for clarity. 

As expected, Figures \ref{fig:PComp}- \ref{fig:P&EComp} show that superlattices with large monoclinic fractions (see Figure \ref{fig:XRDComp}) display low P\textsubscript{r} values and high coercivities (E\textsubscript{c}), or simply exhibit dielectric behaviour (P\textsubscript{r}=0, E\textsubscript{c}=0) in cases where the m-phase is dominant. No antiferroelectric behaviour was observed, confirming that for all superlattices the presence of t-phase ZrO\textsubscript{2} is negligible. The results also show the ferroelectric behaviour of r-phase superlattices differs from that of the equivalent r-phase solid solution. In the case of the HfO\textsubscript{2}-ZrO\textsubscript{2} superlattice family, the best performing superlattice, (HZ\textsubscript{0}-Z)\textsubscript{2}-10,  exhibits similar coercivity and P\textsubscript{r} values more than twice as large ( 33 $\mu$ C/cm\textsuperscript{2}),  
as that of a 10 nm Hf\textsubscript{0.5}Zr\textsubscript{0.5}O\textsubscript{2} monolayer, as shown in Figures \ref{fig:PComp}a and \ref{fig:P&EComp}a. For the (HZ\textsubscript{0}-Z)\textsubscript{n} family, HfO\textsubscript{2}-10 is not shown as it is monoclinic (Figure \ref{fig:XRDComp}) and, therefore, only exhibits dielectric behaviour.

Across Figure \ref{fig:PComp}a-c, the trend for 10 nm Hf\textsubscript{1-x}Zr\textsubscript{x}O\textsubscript{2} solid solution layers shows a non-linear increase in P\textsubscript{r} with increasing Zr content, that saturates for x$>$ 0.75 (see SI). Having such a P\textsubscript{r} increase is consistent with the larger P\textsubscript{r} of r-ZrO\textsubscript{2} when compared to r-HfO\textsubscript{2} reported by DFT calculations \cite{ElBoutaybi2023Electro-opticStudy}. However, Zr content cannot account for the approximately twofold P\textsubscript{r} increase for (HZ\textsubscript{0}-Z)\textsubscript{2}-10 when compared to (HZ\textsubscript{0.5})-10. For this, it must be the case that \(2P_r(\text{HfO}_2-\text{2.5}) + 2P_r(\text{ZrO}_2-\text{2.5}) > P_r (\text{Hf}_{0.5}\text{Zr}_{0.5}\text{O}_2-\text{10})\), \emph{i.e.} that the sum of the polarization of thinner HfO\textsubscript{2} and ZrO\textsubscript{2} sublayers is greater than the polarization of the thicker Hf\textsubscript{0.5}Zr\textsubscript{0.5}O\textsubscript{2} solid solution. 

Instead, it seems that the key to the observed increase in polarization lies in the prevention of strain relaxation such that the strain state of each Hf\textsubscript{1-x}Zr\textsubscript{1}O\textsubscript{2} and ZrO\textsubscript{2} sublayer is maintained in a superlattice construction. This can be seen through the comparison of (HZ\textsubscript{x}-Z)\textsubscript{2}-10 with (HZ\textsubscript{x}-Z)-10 in Figures \ref{fig:XRDComp} and \ref{fig:P&EComp}, where the superlattice with the largest number of sublayers for a given total thickness exhibits better r-phase stability and, consequently, higher P\textsubscript{r} values. In addition, it has been shown that thinner Hf\textsubscript{1-x}Zr\textsubscript{x}O\textsubscript{2} layers exhibit larger polarization values due to a more extended d\textsubscript{111} (the polar axis of the r-phase) \cite{Wei2018AFilms}. It is therefore also possible that through maintaining strain within each sublayer the polarization is further increased due to an additional extension of the d\textsubscript{111}. Therefore, maintaining the strain state within each superlattice sublayer is not only required for stabilization of the ferroelectric phase, but also serves to enhance the polarization of said phase.

Based on the results of HfO\textsubscript{2}-ZrO\textsubscript{2} superlattices, it is then surprising that Hf\textsubscript{0.5}Zr\textsubscript{0.5}O\textsubscript{2}-ZrO\textsubscript{2} superlattices perform noticeably worse despite a lower tendency to relax into the m/t-HfO\textsubscript{2}/ZrO\textsubscript{2} phases. This may be accounted for by a different form of strain accommodation and strain phase diagram for the 50:50 composition, which may cause 2.5 nm layers to be too thin for optimal elongation of the d\textsubscript{111} while 5 nm layers cause too many thickness relaxation effects.

Bringing together the benefits of high Zr content and strain, the Hf\textsubscript{0.25}Zr\textsubscript{0.75}O\textsubscript{2}-ZrO\textsubscript{2} family of superlattices consistently presents the largest P\textsubscript{r} values of all tested compositions, with the best performing (H-Z)\textsubscript{2}-T10 member exhibiting an impressive 42 $\mu$ C/cm\textsuperscript{2} (Figure \ref{fig:P&EComp}).

\begin{figure}[h]
    \centering
    \includegraphics[width=1\textwidth]{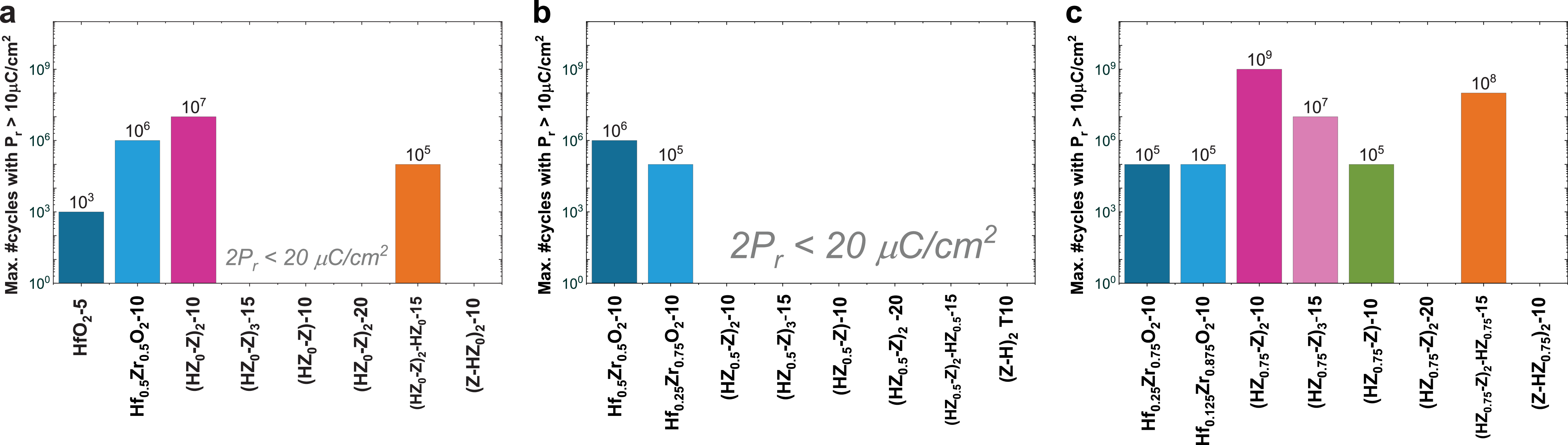}
     \hfill
        \caption{Comparison of the cyclability for superlattices each the three tested families of superlattices (\emph{a)} (HZ\textsubscript{0}-Z)\textsubscript{n}-T, \emph{b)} (HZ\textsubscript{0.5}-Z)\textsubscript{n}-T and \emph{c)} (HZ\textsubscript{0.75}-Z)\textsubscript{n}-T) compared to their solid solution monolayer equivalents. As indicated on the vertical axis, cyclability here is evaluated as the number of cycles for which a memory window of 2P\textsubscript{r}> 20 $\mu$ C/cm\textsuperscript{2} is maintained.}
        \label{fig:Fatigue_Comp}
\end{figure}

For solid solutions there is one big downside to increasing the Zr content: cyclability. Figure \ref{fig:Fatigue_Comp} presents the cyclability of the three superlattice families compared to the monolayers of relevant stoichiometry, where endurance is characterized as the maximum number of cycles for which a 2P\textsubscript{r}>20 $\mu$ C/cm\textsuperscript{2} memory window is preserved. It is shown that with increasing Zr content, solid solution samples can break down after only 10\textsuperscript{4} cycles, despite exhibiting the largest memory window when pristine (2P\textsubscript{r}=37 $\mu$ C/cm\textsuperscript{2}). This is likely because the formation of conductive filaments due to migration of oxygen vacancies upon cycling (the expected breakdown mechanism \cite{Starschich2017PulseHfO2,Xue2018ModelCapacitors}) occurs more readily in ZrO\textsubscript{2}, which has a larger oxygen diffusion coefficient. As a result, when considering solid solutions for applications, the Hf\textsubscript{0.5}Zr\textsubscript{0.5}O\textsubscript{2} composition yields the best compromise of performance (2P\textsubscript{r}=28 $\mu$ C/cm\textsuperscript{2}, Figure \ref{fig:PComp}) and cyclability (10\textsuperscript{6} cycles before breakdown, Figure \ref{fig:Fatigue_Comp}), which is far from optimal.

Fortunately, one of the most consistent differences in the performance of solid solutions and superlattices is that the latter are more cyclable by several orders of magnitude for all tested superlattices. This matches previous observations in ALD-grown HfO\textsubscript{2}-ZrO\textsubscript{2} nanolaminates \cite{Weeks2017EngineeringNanolaminates, Gong2022PhysicalFilm,Lehninger2023FerroelectricReliability} and can likely be attributed to the same mechanism despite the differences between polycrystalline \emph{OIII}-phase nanolaminates and epitaxial r-phase superlattices: accumulation of oxygen vacancies does not only take place at electrode interfaces, but can now also happen at superlattice/nanolaminate interfaces with a similar energy cost. As a result, the density of oxygen vacancies in superlattices with a larger number of sublayers is more homogeneously distributed, which slows down the formation of metallic filaments. Exploiting this property of superlattices while cycling at lower electric fields than those necessary to obtain the maximum P\textsubscript{r}, one can then obtain ferroelectric films that are still highly polar (2P\textsubscript{r}=40 $\mu$ C/cm\textsuperscript{2}) while being cyclable for 10\textsuperscript{9} cycles before breakdown, as shown in the case of the (HZ\textsubscript{0.75}-Z)\textsubscript{2}-10 superlattices.

\section{Conclusions}

This work reports on the successful growth of epitaxial Hf\textsubscript{1-x}Zr\textsubscript{x}O\textsubscript{2}-ZrO\textsubscript{2} superlattices by PLD. Structural characterisation reveals that all constituent sublayers in the superlattices exhibit the polar rhombohedral (\emph{R3m}) phase and, therefore, contribute to the measured ferroelectric polarization, as long as the LSMO-Hf\textsubscript{1-x}Zr\textsubscript{x}O\textsubscript{2} interface is preserved. The largest P\textsubscript{r} values are seen with asymmetric superlattice constructions consisting of thin layers where the strain state in each sublayer is maintained such that d\textsubscript{111} is more elongated.

It was further shown both experimentally and from theory that increasing the Zr content in Hf\textsubscript{1-x}Zr\textsubscript{x}O\textsubscript{2} monolayers facilitates stabilization of the ferroelectric r-phase. Experimentally, it was also seen that increasing Zr content results in larger remnant polarization values at the expense of a worsened device cyclability. In superlattices a similar effect is seen on the polarization, but in this case, the additional interfaces serve to extend the device lifetime. As a result,  superlattice stacks composed of 87.5\% ZrO\textsubscript{2} possess both a large memory window (2P\textsubscript{r}=84 $\mu$ C/cm\textsuperscript{2}) and superb endurance of up to 10\textsuperscript{9} cycles before breakdown (with 2P\textsubscript{r}=40 $\mu$ C/cm\textsuperscript{2}). Since Zr is much more abundant than Hf \cite{2016CRCPhysics}, this form of superlattice construction enables more sustainable "hafnia"-based devices in addition to the established improved device performance.

Despite the differences between the present epitaxial r-phase superlattices and ALD-grown polycrystalline \emph{OIII}-phase nanolaminates, it seems some of the present findings for the former (cyclability, ferroelectric phase stabilization) align with existing research performed on the latter \cite{Weeks2017EngineeringNanolaminates, Migita2021AcceleratedNanolaminates, Lehninger2023FerroelectricReliability}. As a result, trends observed in our work regarding Zr content and rules for the optimal construction of superlattices may be similarly applicable to ALD-grown nanolaminates. If the current findings do translate to other, CMOS compatible, synthesis methods, this method of superlattice construction could serve as an important step towards the fabrication of sustainable and high performance ferroelectric devices.

\pagebreak

\section{Experimental}

All Hf\textsubscript{1-x}Zr\textsubscript{x}O\textsubscript{2} solid solution and Hf\textsubscript{1-x}Zr\textsubscript{x}O\textsubscript{2}-ZrO\textsubscript{2} superlattice samples are grown by PLD on La\textsubscript{0.67}Sr\textsubscript{0.33}MnO\textsubscript{3} (LSMO) buffered (001)-oriented SrTiO\textsubscript{3} (STO) substrates. Polycrystalline Hf\textsubscript{1-x}Zr\textsubscript{x}O\textsubscript{2} targets are synthesised by a solid-state reaction at 1400°C from HfO\textsubscript{2} (99\% purity) and ZrO\textsubscript{2} (99.5\% purity) powders, while the LSMO target was purchased from PI-KEM. A KrF excimer laser with a wavelength of 248 nm is used for ablation. 
LSMO layers with thickness of 30 nm are deposited on STO(001) substrates heated to 775°C, using a laser fluence of 1.6 J/cm\textsuperscript{2} and a frequency of 1 Hz under a 0.15 mbar oxygen atmosphere. Both solid solution monolayers and superlattices are grown subsequently on top of the LSMO layer without cooling or breaking vacuum. The superlattices consist of layers of Hf\textsubscript{1-x}Zr\textsubscript{x}O\textsubscript{2} grown at 800°C using a fluence of 2.0 J/cm\textsuperscript{2} and frequency of 2 Hz under a 0.05 mbar oxygen atmosphere and layers of pure ZrO\textsubscript{2} layers grown under the same conditions but at a substrate temperature of 775°C. Solid solution monolayers are grown under the same conditions as the Hf\textsubscript{1-x}Zr\textsubscript{x}O\textsubscript{2} layers. After deposition, all samples are cooled down to room temperature at a rate of 5°C/min in an oxygen pressure of 300 mbar.

Several variations of Hf\textsubscript{1-x}Zr\textsubscript{x}O\textsubscript{2}-ZrO\textsubscript{2} superlattices are synthesised with three different stoichiometries for the Hf\textsubscript{1-x}Zr\textsubscript{x}O\textsubscript{2} layer (referred to as families in this text): HfO\textsubscript{2}, Hf\textsubscript{0.50}Zr\textsubscript{0.50}O\textsubscript{2} and Hf\textsubscript{0.25}Zr\textsubscript{0.75}O\textsubscript{2}. (HZ\textsubscript{x}-Z)\textsubscript{n}-T is used to refer to each superlattice variation within a family, where HZ\textsubscript{x}-Z refers to a superlattice starting with a Hf\textsubscript{1-x}Zr\textsubscript{x}O\textsubscript{2} layer (Z-HZ\textsubscript{x} then represents those starting with a ZrO\textsubscript{2} layer), \emph{n} is the number of superlattice HZ\textsubscript{x}-Z repetitions and \emph{T} is the total thickness of the superlattice in nm. The layer thickness ratio between the two sublayers in the superlattice period is kept as 1:1 unless otherwise stated. The impact of the number of repetitions, layer thickness, stack symmetry and stack order are verified for each family. For monolayers, the superlattice notation is not used, however the total thickness (T) of the layer is indicated next to the composition as Hf\textsubscript{1-x}Zr\textsubscript{x}O\textsubscript{2}-T.

Structural characterisation was primarily performed by x-ray diffraction using a Panalytical X'Pert Pro diffractometer in line focus for $\theta-2\theta \ $ scans and in point focus for pole figures. Additional in-plane measurements around poles were performed using a Bruker D8 Advance diffractometer equipped with a TXS (Turbo X-ray Source) rotating anode and an Eiger2 R 500k area detector.

A cross-section TEM lamella was made using focussed ion beam (FIB) milling with a Ga\textsuperscript{+} ion beam in the (100) zone of the STO substrate using an FEI Helios G4 CX FIB-SEM. The lamella was Ga\textsuperscript{+} ion polished at 5 kV and 2 kV to a thickness of approximately 100 nm and plasma cleaned in 25/75 O\textsubscript{2}/Ar atmosphere for 1 minute just before imaging. 

Scanning transmission electron microscopy (STEM) images were collected using a double-aberration corrected ThermoFisher Themis Z (S)TEM microscope operated at 300 kV with a beam current of 50 pA and a convergence semiangle of 25 mrad. A high-angle annular dark field image was obtained with collection angles of 27-164 mrad.  
For electrical characterization, circular Ti(4 nm)/Au(40 nm) top electrodes of various sizes are patterned by photolithography and deposited by e-beam evaporation. Ferroelectric characterization is performed using an aixACCT TF Analyser 2000, where the remnant polarization is characterized on pristine films with 1 kHz PUND measurements \cite{Scott1988SwitchingMemories} and cyclability is tested by fatigue measurements (100 kHz, 1 kHz PUND).

The density functional theory (DFT) calculations were performed using a plane-wave basis set as implemented in the Vienna ab-initio simulation package (VASP) \cite{Kresse1996EfficientSet, Kresse1996EfficiencySet}. The Perdew-Burke-Ernzerhof (PBE) form of the generalized gradient approximation (GGA) with the PBEsol modification is employed to approximate the exchange-correlation functional \cite{Perdew2008RestoringSurfaces}. The plane-waves are expanded up to a cutoff of 600 eV. The following valence states are explicitly considered for the different elements: O – 2s$^2$, 2p$^4$; Hf – 5p$^6$, 5d$^2$, 6s$^2$; Zr – 4s$^2$, 4p$^6$, 4d$^2$, 5s$^2$. For (111)-oriented pure HfO$_2$ and ZrO$_2$, 36 atom simulation cells were used, with a 4x4x2 k-mesh was used to sample the irreducible Brillouin zone and a proportional reduction to 4x4x1 for the HfO$_2$/ZrO$_2$ superlattices, with 72 atoms in the simulation cell. The total cohesive energy and the Hellmann-Feynman forces on each atom were relaxed below 10$^{-6}$ eV and 0.01 eV/Å respectively, for all structures. For HfO$_2$-ZrO$_2$ superlattices with equally thick HfO$_2$ and ZrO$_2$ layers, the formation energy per cation is defined as \cite{Mukherjee2024First-principlesSuperlattices},

\begin{equation}
    \Delta E_{Hf/Zr}^{SL} = E_{Hf/Zr}^{SL} - \frac{1}{2}(E^{Bulk}_{gs-HfO_2} + E^{Bulk}_{gs-ZrO_2})
\end{equation}
where \(E^{SL}_{Hf/Zr}\) is the cohesive energy per cation of the superlattice, and \(E^{Bulk}_{gs-HfO_2}\) and \(E^{Bulk}_{gs-ZrO_2}\) are the cohesive energies per cation of the ground states of HfO$_2$ and ZrO$_2$, respectively.

\section{Acknowledgments}
The authors thank Jacob Baas, Henk Bonder and Joost Zoestbergen for technical support. We thank Majid Ahmadi for help with electron microscopy and lamella preparation. This publication is part of the TRICOLOR project, which is financed by the Dutch Research Council (NWO) through grant OCENW.M20.005 and by the Luxembourg National Research Fund (FNR) through grant INTER/NOW/20/15079143/TRICOLOR.

\section{References}
\begin{multicols}{2}
    \singlespacing
\printbibliography[heading=none]
\end{multicols}
\pagebreak

\section{Supplementary Information}
\setcounter{figure}{0}
\renewcommand{\figurename}{Figure}

\begin{figure}[ht]
    \centering
    \includegraphics[width=0.6\textwidth]{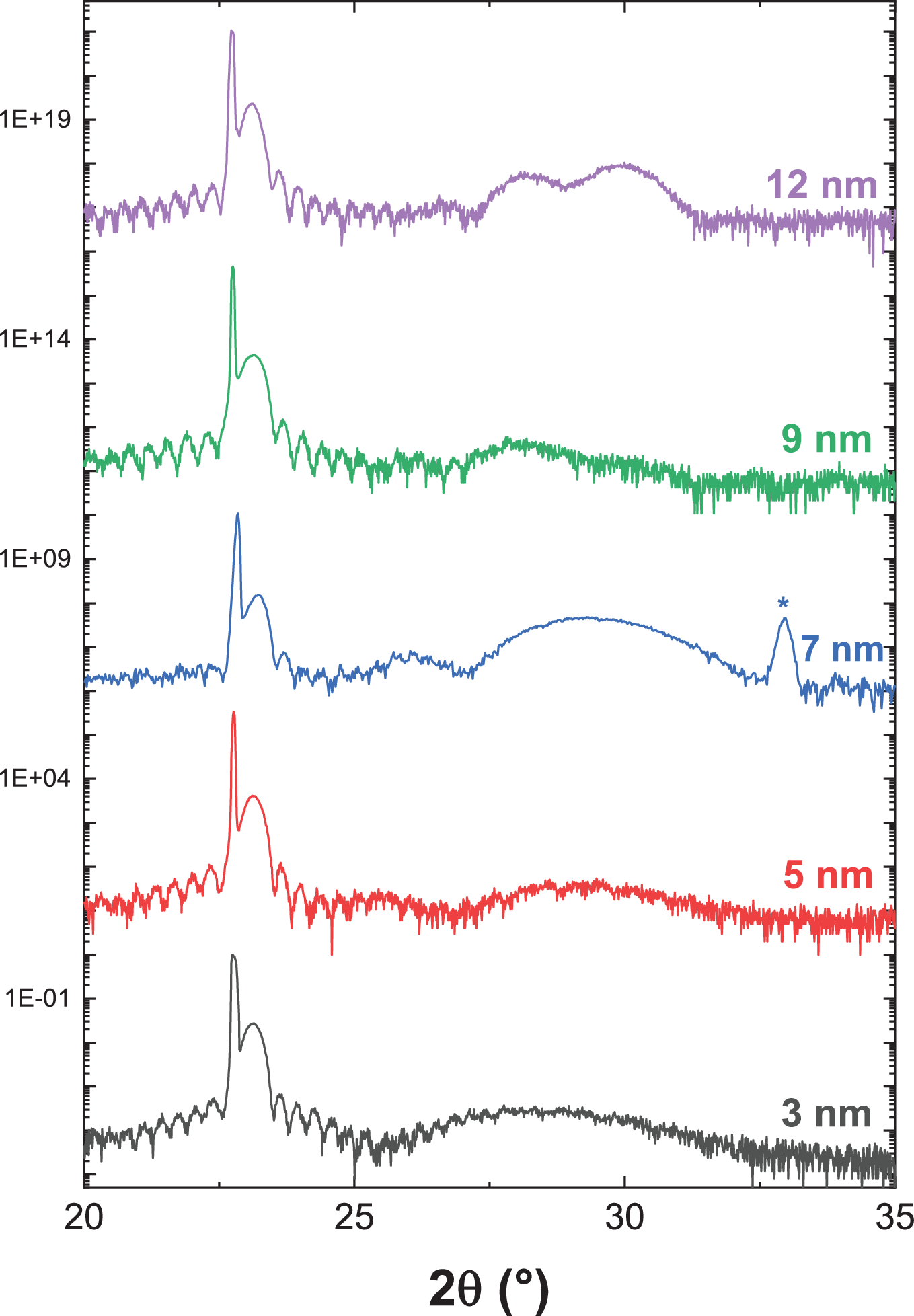}
     \hfill
        \caption{\(\theta/2\theta\) measurements for epitaxial HfO$_2$ thin films grown on STO(100)/LSMO of different thicknesses. With increasing thickness the d$_{111}$ spacing of the r-phase decreases until the films begin to relax into the monoclinic phase (observed at $\sim$28° above 7 nm film thickness).}
        \label{fig:ThicknessRelaxation}
\end{figure}

\begin{figure}[ht]
    \centering
    \includegraphics[width=0.6\textwidth]{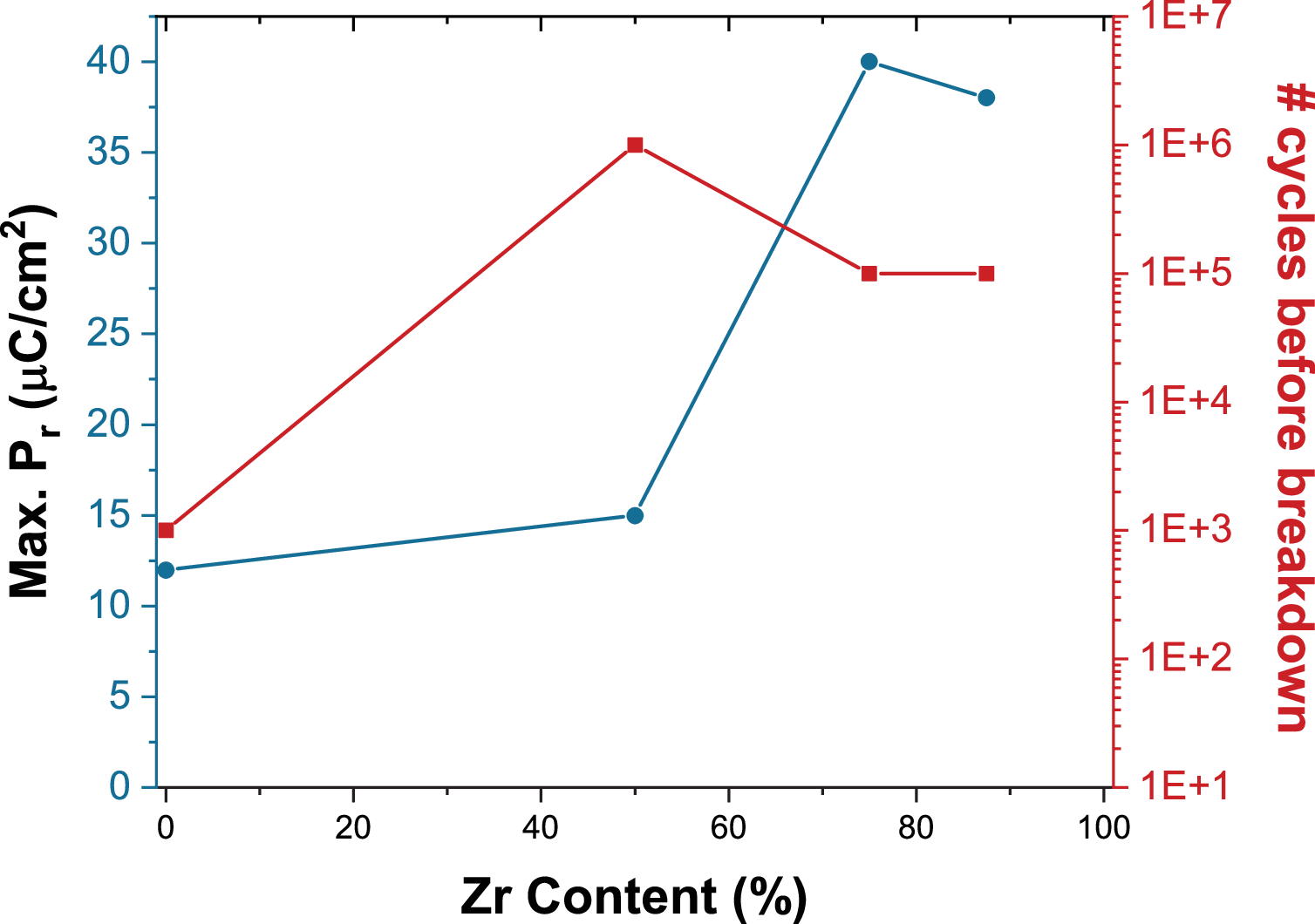}
     \hfill
        \caption{Comparison of the maximum remanent polarization and cyclability for 10 nm thick Hf$_{1-x}$Zr$_x$O$_2$ solid solution monolayers as a function of Zr content.}
        \label{fig:Zrcontent}
\end{figure}

\begin{figure}[ht]
    \centering
    \includegraphics[width=1\textwidth]{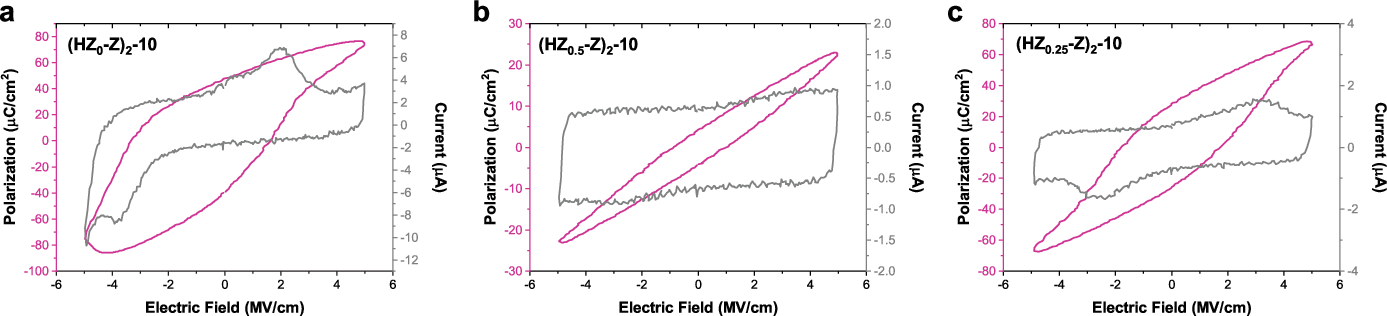}
     \hfill
        \caption{Dynamic hysteresis measurements corresponding to the best performing superlattices of each Hf$_{1-x}$Zr$_x$O$_2$ composition shown in Figure 4 of the main text.}
        \label{fig:DHM_SL}
\end{figure}

\begin{figure}[ht]
    \centering
    \includegraphics[width=0.6\textwidth]{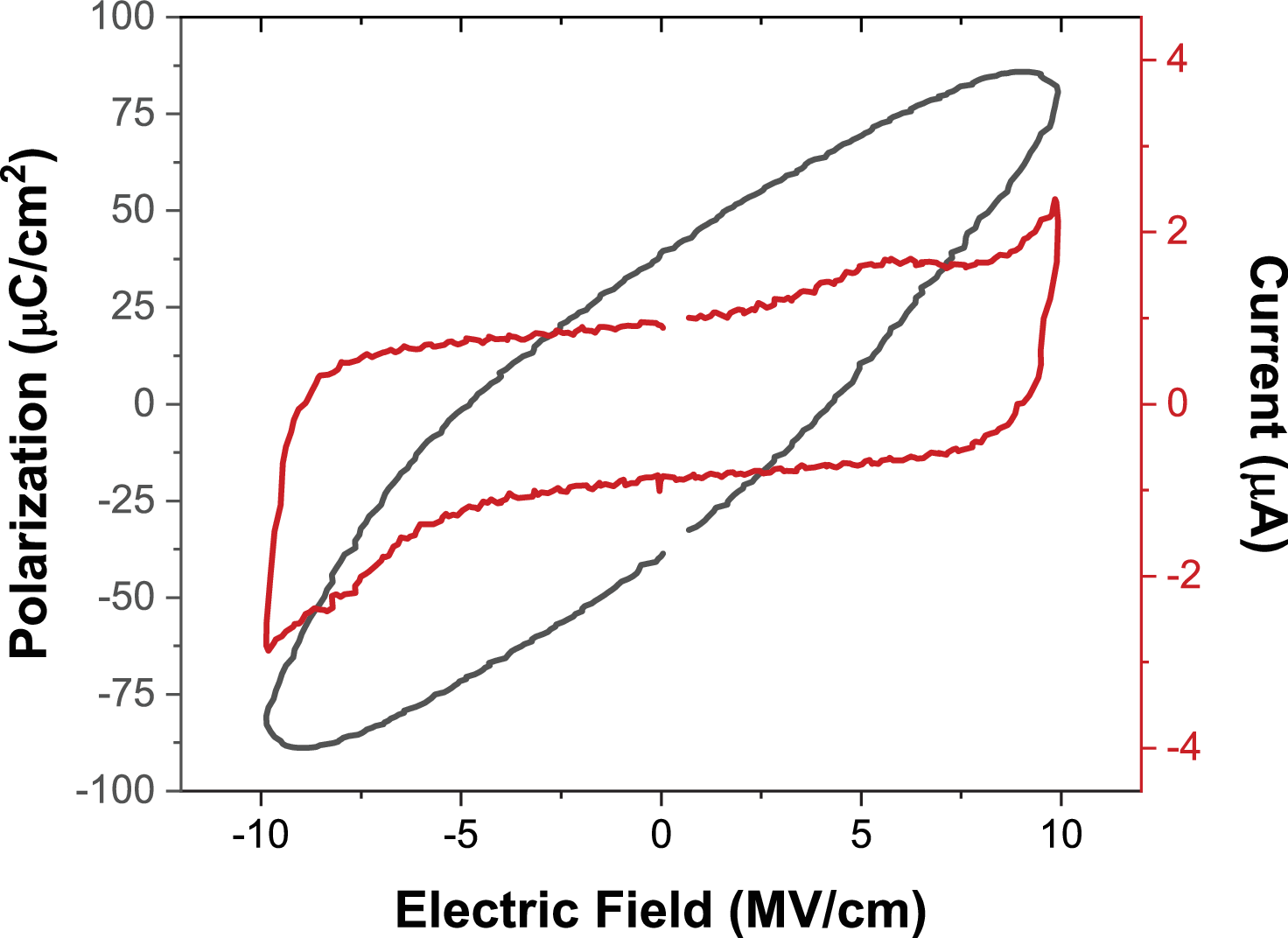}
     \hfill
        \caption{Dynamic hysteresis measurement of a 5 nm HfO$_2$ monolayer. The remanent polarization extracted from PUND is 10 $\mu$ C/cm$^2$.}
        \label{fig:DHM_HfO2}
\end{figure}

\begin{figure}[ht!]
    \centering
    \includegraphics[width=0.9\textwidth]{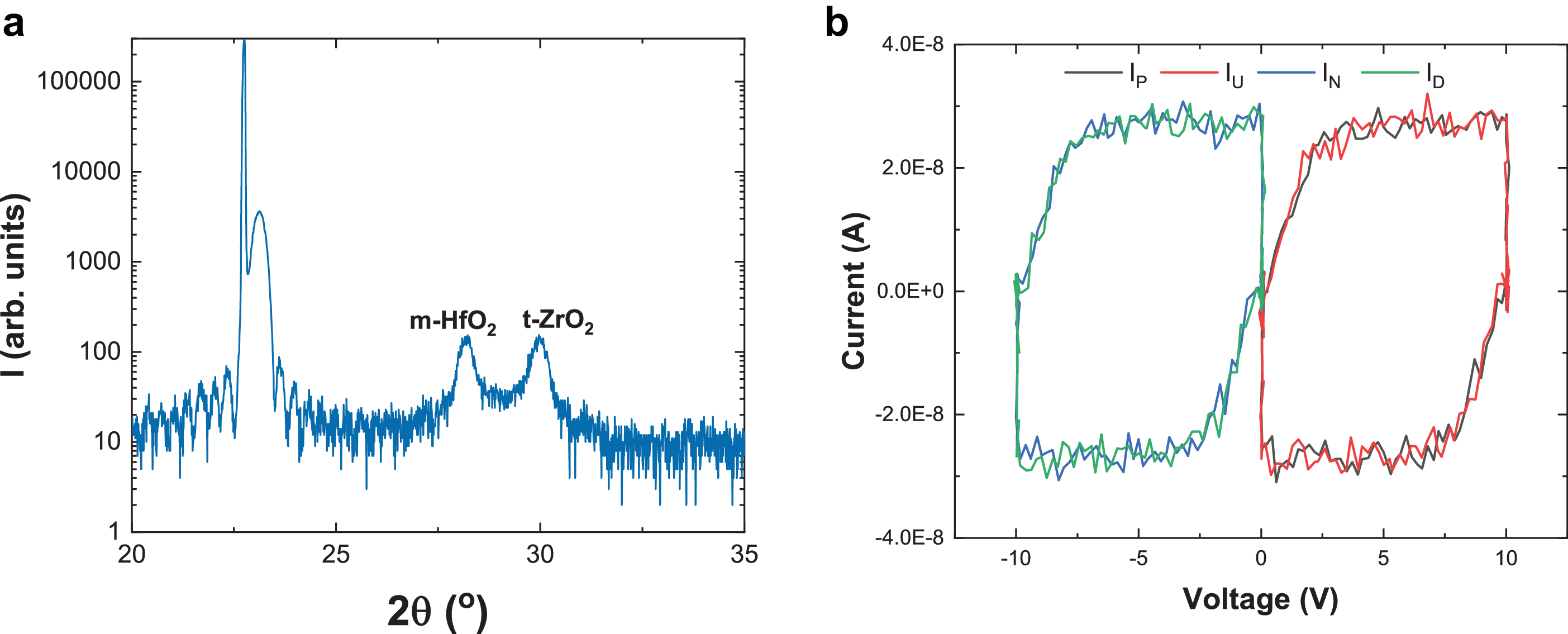}
     \hfill
        \caption{\emph{a)} \(\theta/2\theta\) measurement and \emph{b)} PUND current loops for a (HZ$_0$-Z)$_2$-24 superlattice. When the sublayer thickness exceeds 5 nm both the HfO$_2$ and ZrO$_2$ layers relax into lower energy phases that exhibit purely dielectric behavior.}
        \label{fig:TooThickSL}
\end{figure}

\end{document}